\documentclass[letter]{aa} 
\usepackage{txfonts}
\usepackage{graphicx,epsfig,fancyhdr,rotating,amsmath,natbib}

\begin{document}

   \title{$H\alpha$ observations of the Be/X-ray binary MWC 656
    } 
   \titlerunning{The Be/X-ray binary MWC 656}
   \author{R. Zamanov\inst{1}\fnmsep\thanks{\email{rzamanov@nao-rozhen.org}}
          \and  V. Marchev\inst{1} 
          \and  K. A. Stoyanov\inst{1}
          \and  M. D. Christova\inst{2}
          \and  M. Minev\inst{1}
	  \and  M. Moyseev\inst{1} 
	  \and  D. Marchev\inst{3}
          \and  J. Mart\'i\inst{4}
          \and  M. F. Bode\inst{5,6}
          \and  P. Markishki\inst{1}
          \and  S. Stefanov\inst{1}
	  }
 \institute{Institute of Astronomy and National Astronomical Observatory, 
            Bulgarian Academy of Sciences, 
            Tsarigradsko Shose 72, BG-1784 Sofia, Bulgaria  
         \and 
         Department of Applied Physics, Technical University of Sofia, 
	 8 Kliment Ohridski blvd., 1000 Sofia, Bulgaria
      \and 
       Department of Physics and Astronomy, Shumen University "Episkop Konstantin Preslavski",  
       115 Universitetska Str., 9700 Shumen, Bulgaria
         \and 
	Departamento de F\'isica, Escuela Polit\'ecnica Superior de Ja\'en, 
	Universidad de Ja\'en, Campus Las Lagunillas, A3, 23071, Ja\'en, Spain 	 
         \and
      Astrophysics Research Institute, Liverpool John Moores University, IC2, 149 Brownlow Hill, Liverpool, L3 5RF, UK
         \and
      Office of the Vice Chancellor, Botswana International University of Science and Technology, 
      Private Bag 16, Palapye, Botswana         
       % \thanks{The institute accepts e-mails}
             }

   \date{Received June 15, 2026; accepted August 25, 2026}

\abstract 
{MWC~656 is a Be/X-ray binary identified as an emission-line object,
X-ray and radio source
and might be a weak $\gamma$-ray source, 
harbouring an enigmatic secondary component.  }
{ Our aim here is to study the orbital modulation of the H$\alpha$ emission line, disc truncation 
and to evaluate the binary parameters. }
{During the period 2021-2026, we obtained high resolution ($\lambda / \Delta \lambda \approx 30000$) 
optical spectra of MWC~656 and analysed them together with the published data. 
}
{ We find an orbital modulation of the $H\alpha$ emission line with period 
$P_{orb} = 60.43 \pm 0.03$~d. 
By assuming that the Be circumstellar disc is truncated by the 
companion, we estimate orbital eccentricity $e=0.11 \pm 0.01$,
semimajor axis $a \approx 160$~R$_\odot$, and mass of the binary $M_1 + M_2 \approx  15.1$~M$_\odot$. 
}  
{ 
The orbital modulation of the $H\alpha$ emission suggests that 
(1) in this object we have a new type of cyclic variability of a Be disc; 
(2) the orbit is mildly eccentric;
and 
(3) the secondary component of MWC~656  is a massive object $\approx 5$~M$_\odot$, 
more massive than the neutron stars -- a stellar-mass black hole or 
a massive stripped star. 
}

   \keywords{ Stars: emission-line, Be -- X-rays: binaries --
               accretion, accretion discs -- stars: individual: MWC~656   }

   \maketitle

%\nolinenumbers

\section{Introduction}

MWC~656 (HD 215227) is a 9th magnitude star in the constellation Lacerta identified as an
emission line object in the Mount Wilson Observatory surveys (Merrill \& Burwell 1933). 
In July 2010 the $AGILE$ satellite detected a $\gamma$-ray transient source, 
AGL J2241+4454 in its direction (Lucarelli et al. 2010; 
Munar-Adrover et al. 2016 and references therein), 
that  activated multi-wavelengths observations. 
The primary is a rapidly rotating Be star that forms 
an outwardly diffusing gaseous disc. It has projected
rotational velocity  $v \sin i \sim 300$~km~s$^{-1}$  (M{\"u}ller-Horn et al. 2026), 
and is classified as B1.5~III (Casares et al. 2014). 
MWC~656 is the first binary system identified to contain a Be star and a black hole
(Casares et al. 2014). 
However some recent studies argued  that the secondary could be 
of a different type -- 
a neutron star, a white dwarf, a sdO star or a stripped He star
(Janssens et al. 2023,  Rivinius et al. 2024, 
Dzib \& Jaron  2025, M{\"u}ller-Horn et al. 2026). 

The Be/X-ray binaries  are systems that consist of a compact
object orbiting an optical companion that is an Be star. Be stars
are non-supergiant fast-rotating O-type and B-type and luminosity class
III-V stars which, at some point of their lives, have shown spectral lines in
emission (Slettebak 1988; Porter \& Rivinius 2003, Rivinius et al. 2013).  
The best studied lines are those of hydrogen (Balmer and Paschen series), but the
Be stars may also show He and Fe in emission (see Hanuschik 1996, and
references therein). They also typically display an amount of infrared excess.
The origin of the emission lines and infrared excesses is attributed
to an equatorial disc, fed from material expelled from the rapidly rotating
Be star. In the  Be/X-ray binaries the compact object interacts with the 
Be star circumstellar disc, truncates it, and accretes material from it (Reig 2011). 

Here, we present new spectra of MWC~656 obtained during 2021-2026, 
and combine  them with the published data
to estimate some of the critical parameters of the binary system.

 \begin{figure}     
  \vspace{11.0cm}   
  \includegraphics{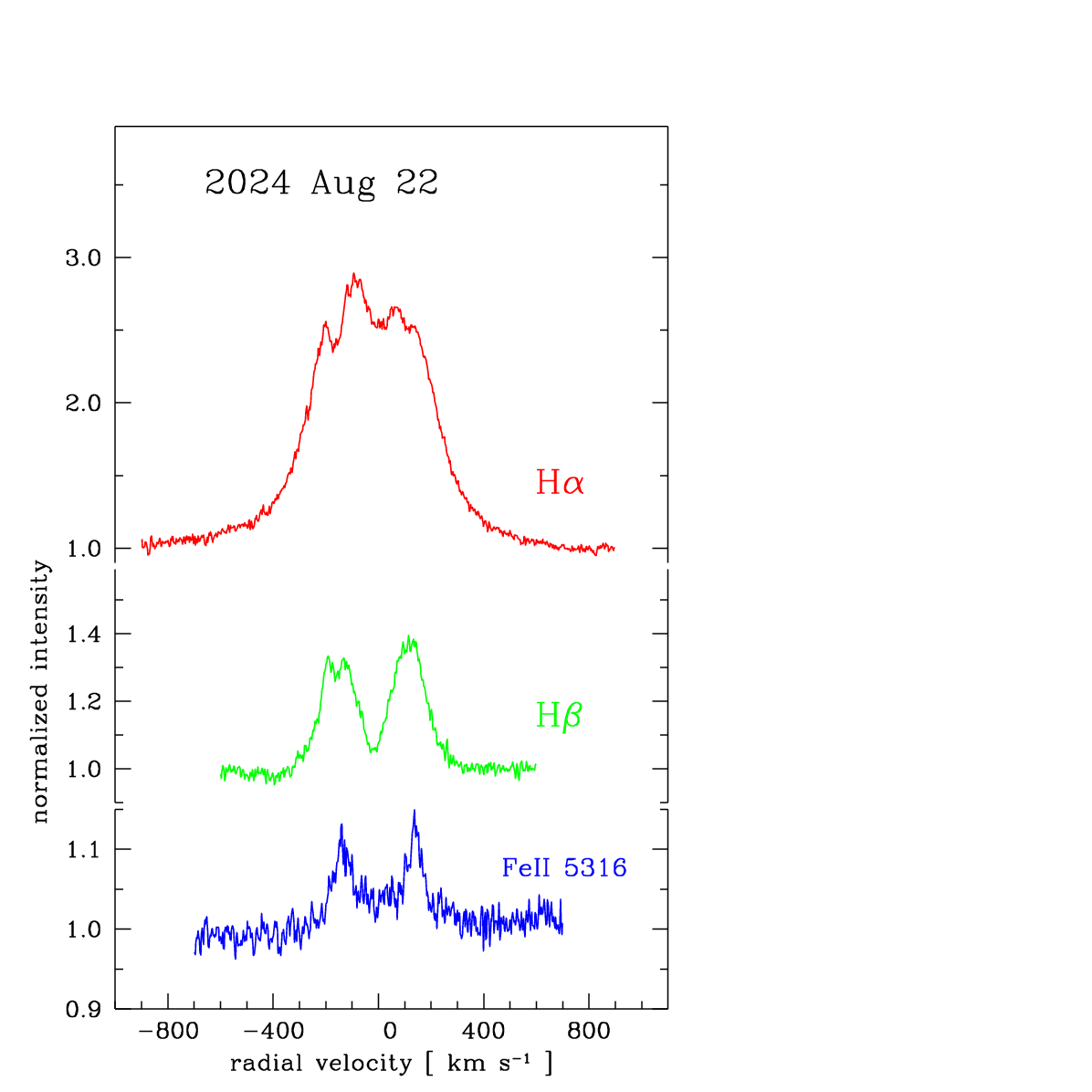} 
  %\special{psfile=MWC656pdm.eps    hoffset=250  voffset=-70  hscale=50  vscale=50  angle=0} 
  \caption[]{Prominent emission lines in the spectrum of MWC~656.    
  	    } 
  \label{f.1}  	   
\end{figure}	     

\begin{figure*}     
  \vspace{8.3cm}   
  \includegraphics{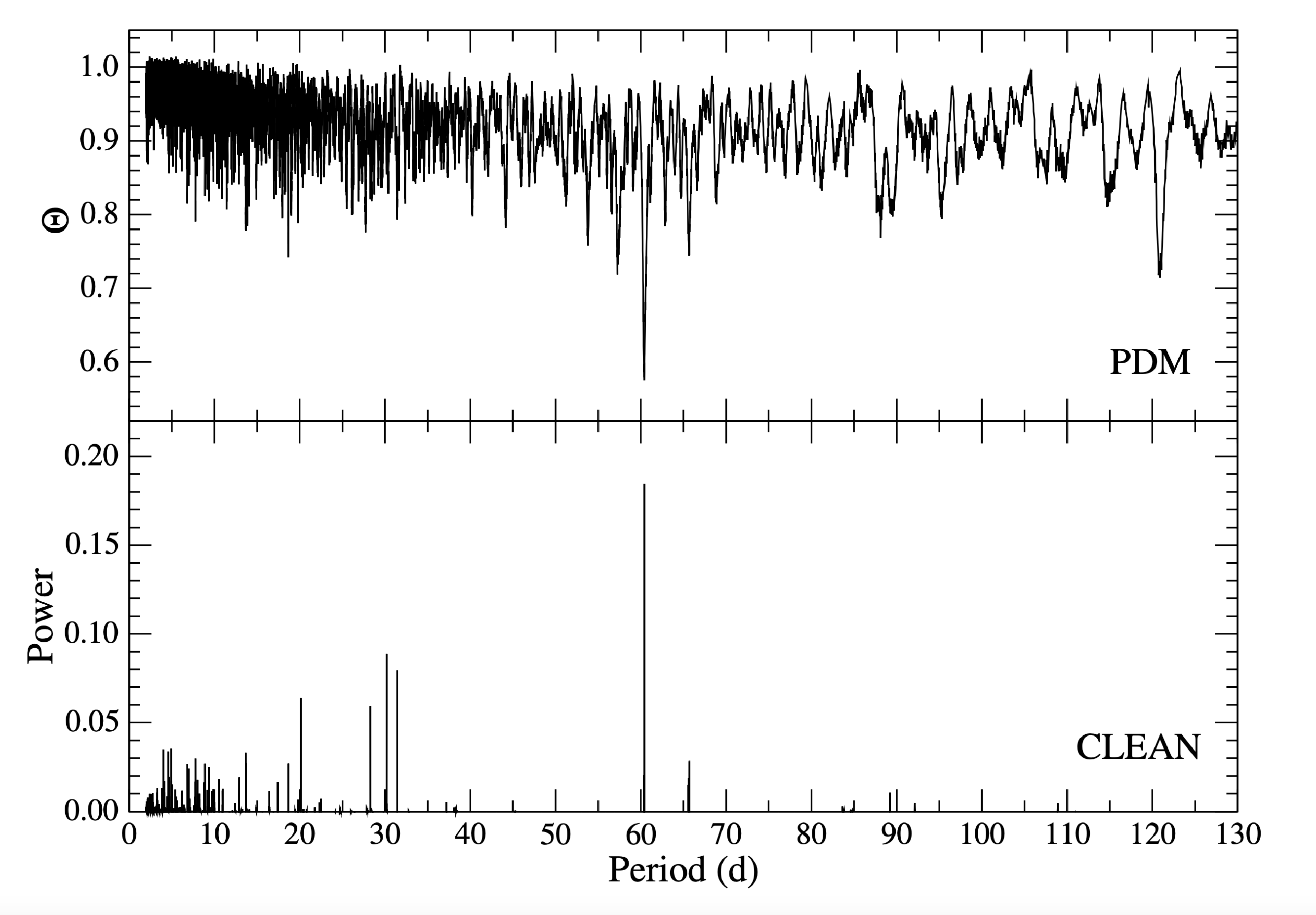} 
  \includegraphics{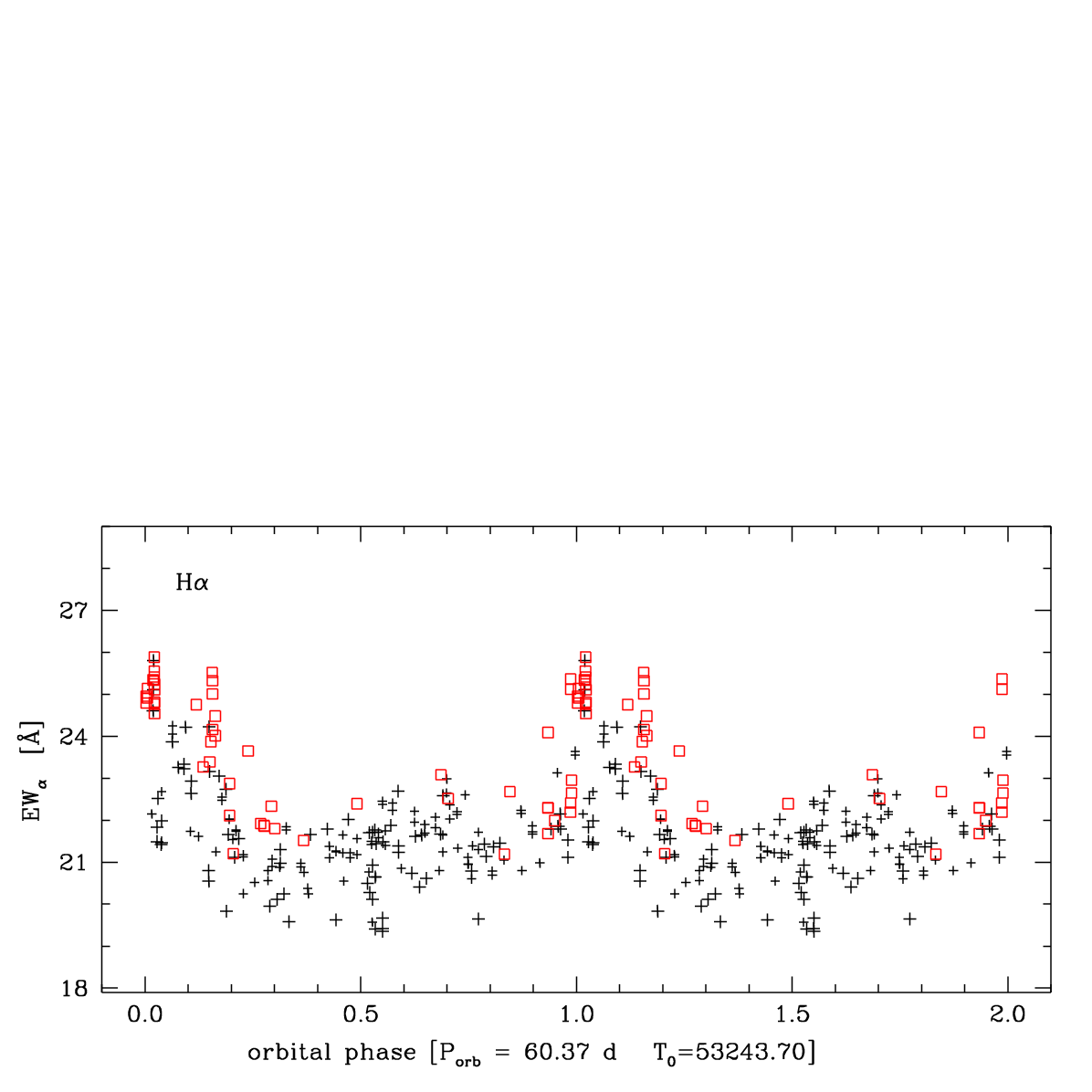} 
  \caption[]{The left panel represents periodograms for the EW$_\alpha$. They indicate 
    a period of $60.43 \pm 0.03$~d.
    The right panel represents the orbital modulation of the EW$_\alpha$.   
    The folded curve is shown twice for clarity. 
    The black plusses indicate the published data, 
    the red squares, the new measurements.  
      } 
  \label{f.orb}  	   
  \vspace{8.3cm}   
  \includegraphics{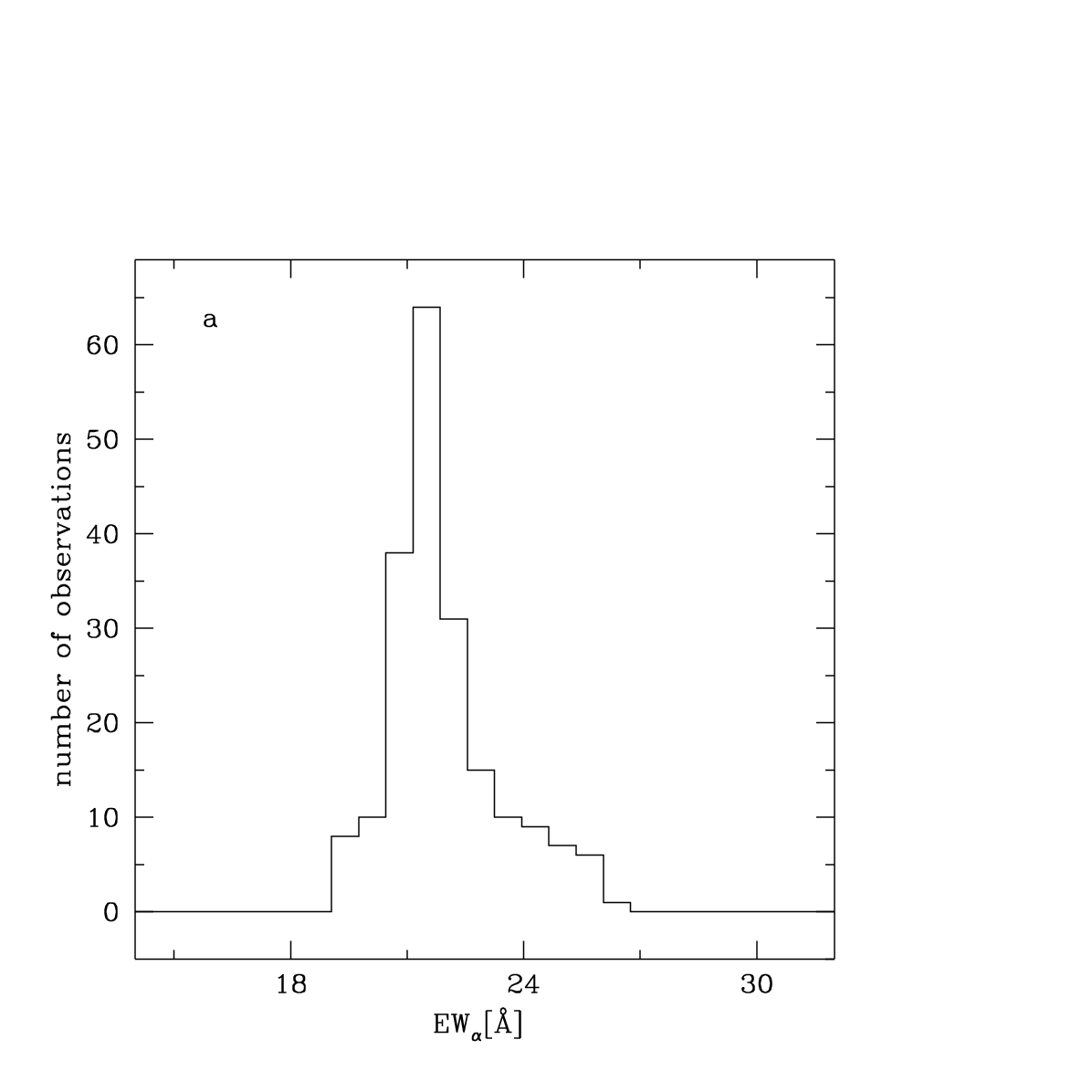} 
  \includegraphics{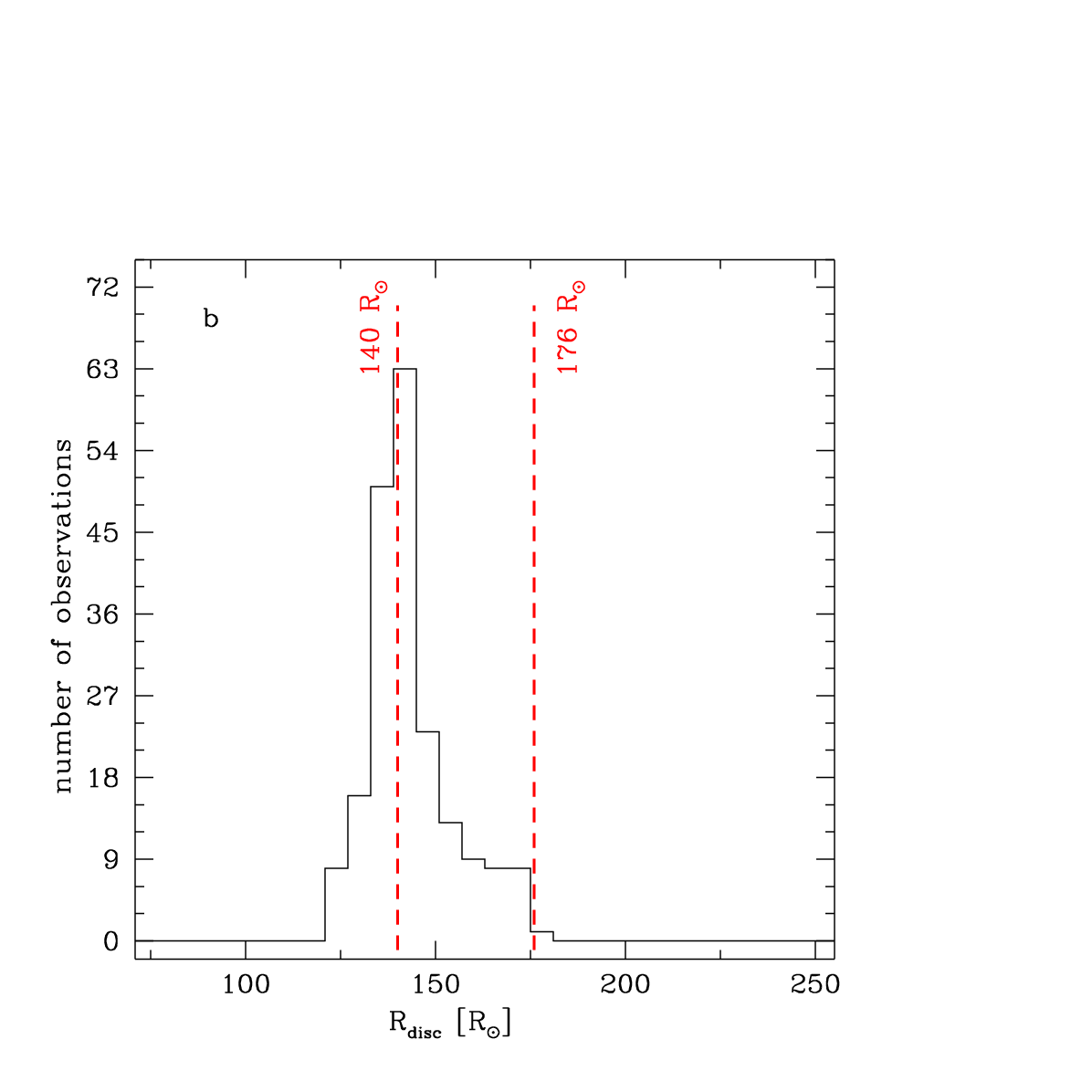}  
  \caption[]{Histograms of the EW$_\alpha$ and of the disc radius. 
             The vertical dashed lines indicate the disc truncation 
	     by the orbit of the companion.  }
  \label{f.hist}  
\end{figure*}

\section{Observations}
\label{s.obs}

During the period  from November 2021 to March 2026,
we secured 53 optical spectra of MWC~656 with 
the ESpeRo Echelle spectrograph (Bonev et al. 2017) 
on the 2.0 m RCC telescope of the Rozhen  National Astronomical Observatory, Bulgaria. 
An example of the prominent emission lines on our spectra is plotted in Fig.~\ref{f.1}, where 
the X-axis is heliocentric radial velocities, the Y-axis is 
intensity normalized to the local continuum.   
On each spectrum, we measure  the equivalent width of the  H$\alpha$ emission line (EW$_\alpha$).
The measurements are summarized in Table~A.1. The typical error
of  EW$_\alpha$ is $\pm 0.5$~\AA.  

\section{Results}

\subsection{Periodogram analysis}
\label{s.pa} 

Combining published (Zamanov et al. 2022)
and new data we have 218 measurements of the EW$_\alpha$
for the period April 2011 until  April 2026. 
For periodogram analysis, we used  the Phase Dispersion Minimization (PDM) (Stellingwerf 1978) 
and the CLEAN algorithm (Roberts et al. 1987).
The periodograms for the EW$_\alpha$ of MWC~656 are shown in Fig.~\ref{f.orb}.
A period of 60.43~d is independently detected in both cases as the most significant one. 
A zoom-in of the results is shown in Fig.~\ref{f.zoom} in the Appendix. 
MWC~656 shows a photometric periodicity of $60.37 \pm 0.04$ days  with amplitude 
0.05 magnitudes (Williams et al. 2010). Our periodogram analysis
of the $EW_\alpha$ gives very similar value $60.43 \pm 0.03$~days.
These results  strongly suggest that the orbital period
is in the range $60.25 < P_{orb} < 60.52$~d. The orbital variability 
of EW$_\alpha$ is plotted in Fig.~\ref{f.orb}. 
The black plusses indicate the published data, 
the red squares, the new measurements. 

\subsection{Parameters of the primary}
\label{s.Be}

The inclination of the Be star is connected with  the full width at zero intensity of the FeII lines and its radius:  
\begin{equation}
   \frac{\rm FWZI}{2 \sin i }=\left(\frac{\rm G \; M_1}{ \rm R_1}\right)^{1/2},   \\  %  ^ {\frac{1}{2}}   \\
\label{eq.FWZI}
\end{equation}
where G is the gravitational constant, M$_1$ is the mass of the Be star, 
R$_1$ is its radius, $i$ is the inclination of the Be star to the line of sight.
Eq.~\ref{eq.FWZI} is identical to that used in Sect.~6.1 of Casares et al. (2012).
It represents the Keplerian motion in the disc 
and that the region where the Fe II lines are produced can be extended down 
to the surface of the Be star. 
%The parameter $\eta$, for which we adopt $ 0 \le \eta < 0.1$,
%represents how close to the surface of star the emission at FWZI of the FeII lines is formed.
The rotational period of the Be star is also connected with the above parameters: 
\begin{equation}
   {\rm P_{\rm rot}} = \frac { 2 \pi R_1} { v \sin i } \sin i ,
\label{eq.Pr}
\end{equation}
where $v \sin i $ is the projected rotational velocity of the primary.  
The radial velocities of the spectral lines (see Table~1 in  M{\"u}ller-Horn et al. 2026)
give: 
\begin{equation}
   {\rm M_1} \sin ^3 i = 5.13 \pm 0.47 \; M_\odot.
\label{eq.3}
\end{equation}
In the above equations we adopt:   
the rotational period of the primary
$P_{rot} = 1.12 \pm 0.03$~d visible in the TESS photometry,
$v \sin i = 313 \pm  3$~km~s$^{-1}$, and 
$FWZI = 710 \pm 20$~km~s$^{-1}$ (Zamanov et al. 2021).  

Solving numerically the above equations we find most probable values:  
$R_1 = 9.1 \pm 0.1$~R$_\odot$,  
$M_1 = 9.0-10.6$~M$_\odot$, 
and $i \approx 50 - 53^\circ$. 
To achieve a "perfect" agreement with the measured values, 
we have to accept a small misalignment 
between the inclination of the Be star ($\approx 50^\circ$)
and the orbital inclination ($\approx  53 ^\circ$).
The uncertainties in the equations are discussed in Appendix.~\ref{s.a2}. 

\subsection{Radius of the circumstellar disc}

Having in mind that the discs of the Be stars are near Keplerian (Porter \& Rivinius 2003, Meilland et al. 2012) and that the Be stars rotate at rates below the critical rate (e.g. Chauville et al. 2001; Zorec et al. 2017),  
we calculate the disc radius by the following  formula (Zamanov et al. 2016):
%-----------------------------------------------------
 \begin{equation}
        \frac {R_{disc}}{R_1}  = \; (1 -\epsilon) \; 0.467 \; \; EW_\alpha^{1.183}, 
  \label{e.Rd}
  \end{equation}
%------------------------------------------------------------
where $\epsilon$ is a dimensionless parameter, for which  we adopt $\epsilon \approx 0.1$.
$R_1$ is radius of the primary,  $R_1 = 9.1 \pm 0.1$~R$_\odot$. 
This equation  is  similar to, but slightly different from,
that used by Coe et al. (2006) and Monageng et al. (2017)
and expresses the fact that $R_{disc}$ 
grows as EW$_\alpha$ becomes larger (e.g. Grundstrom \& Gies 2006).

\subsection{Orbital eccentricity}
The  EW$_\alpha$ of MWC~656 has an average value of $22.2 \pm 1.5$~\AA\ 
median value 21.8~\AA, and varies in the range from 19.4 to 25.9 \AA. 
The maximum of $EW_\alpha$ is  around the periastron passage of the companion, 
when the EW$_\alpha$ is about 3~\AA\ larger. 
We note that the spectral observations of  PSR B1259-63/LS 2883
(a Be/X-ray binary containing a neutron star with a well known orbit)
 during four periastron passages 
(Chernyakova et al. 2020, 2021,  van Soelen et al. 2016)
 reveal an increase
of the $EW_\alpha$ at the time of the periastron of the neutron star.

In MWC~656, when the companion is at apastron  $EW_\alpha \approx 21.3$~\AA, while when 
at periastron  $\approx 24.5$~\AA.
Following Eq.~\ref{e.Rd}, we estimate that disc size increases 
from 138~R$_\odot$ to 163~R$_\odot$.
This implies that the H$\alpha$ disc pulsates each orbital period
with a semi-amplitude of about 15~R$_\odot$. 

In Fig.~\ref{f.hist}a  we plot the histogram of the $EW_\alpha$. The 
histogram has one well-defined peak at $EW_\alpha = 21.3$~\AA. 
The tendency for the disc emission fluxes to cluster at specified levels 
is related to the truncation of the disc at specific
disc radii by the orbiting compact object (e.g. Coe et al. 2006). 
In binary stars, disc truncation is a phenomenon 
in accretion discs (e.g. Hameury \& Lasota 2006) 
as well as in outflowing discs (Okazaki \& Hayasaki 2007). 
In  the Be/X-ray binaries the limiting radii are defined 
by the closest approach of the companion
(Okazaki \& Negueruela 2001). Following this, we adopt that 
the peak of the distribution corresponds to the closest approach of the companion:
  \begin{equation}
    a (1-e) = 140 \; R_\odot, 
  \label{eq.e1}
  \end{equation}
and that the maximum size of the disc is limited by the
the maximum distance between the components:
  \begin{equation}
   a (1+e) = 176 \; R_\odot. 
  \label{eq.e2}
  \end{equation}
From Eqs.~\ref{eq.e1} and \ref{eq.e2} we derive  
semimajor axis $a= 160$~R$_\odot$ and eccentricity $e=0.11$.

\subsection{Masses of the components}

The semi-major axis $(a)$ is related to the orbital period $(P_{orb})$ by Kepler's third law:
$  P_{orb}^2 = 4 \; \pi^2 \; G^{-1} \; a^3  (M_1 + M_2)^{-1}  $ , 
%\label{eq.K3}
where $G$ is the gravitational constant, 
$M_1$ is the mass of the  primary, and $M_2$ is the mass of the secondary. 
Using $P_{orb}=60.4$~days and $a= 160$~R$_\odot$, 
we evaluate  $M_1 + M_2 = 15.1$~M$_\odot$. 
If $M_1= 9.0 - 10.6$~M$_\odot$ (see Sect.\ref{s.Be}),  
we find  $M_2 \approx 4.5 - 6.1$~M$_\odot$.

\section{Discussion}

The obtained value of $R_1 \approx 9.1$~R$_\odot$ is similar to 
the average radius for a B1-2 III star of  
9.3 -- 8.3~R$_\odot$ (Straizys \& Kuriliene 1981).
Hohle et al. (2010) estimate that 
the average mass of B1~III stars is $11.98 \pm 1.7$ and
of B2~III --  $7.94 \pm 1.0$~M$_\odot$.
The value $M_1= 9.0 - 10.6$~M$_\odot$ found in Sect~\ref{s.Be} 
is in agreement with the  average mass of B1.5~III stars. 

For MWC~656, Casares et al. (2014) suggested that it contains a Be star and a black hole.  
Additional evidence for the black hole nature of the secondary 
component came from the faint X-ray luminosity of the system
$L_X \approx 3 \times 10^{30}$~erg~s$^{-1}$ 
and its location in the radio/X-ray luminosity plane, which is consistent with the
properties of quiescent black hole binaries (Dzib et al. 2015; Ribo et al. 2017).

In the last few years,  Janssens et al. (2023), Rivinius et al. (2024) 
and M{\"u}ller-Horn et al.(2026) cast doubts 
as to whether the secondary is a black hole, and proposed that it is an sdO subdwarf,
a neutron star, a white dwarf, or a massive stripped star. 
We note that most sdO stars have masses 
in the range 0.40 -- 0.55~M$_\odot$ (Zhang, Chen \& Han 2010), 
lower than 
the masses of the neutron stars in Be/X-ray binaries 1 -- 2.4~M$_\odot$ (Kaper et al. 2006). 
Our result indicates that the secondary component of MWC~656 
is in fact more massive than the neutron stars in the Be/X-ray binaries. 
The possible nature could be either a 
black hole (Casares et al. 2014; Ribo et al. 2017) or a massive stripped star
(M{\"u}ller-Horn et al. 2026). Several thousands of such stripped helium stars with masses of 1 -- 7~M$_\odot$ are expected to exist in the Milky Way but none of
them have been identified so far (Yungelson et al. 2024).

The quasi-periodic variations  of circumstellar discs of the  Be  stars
are interpreted as due to an eccentric disc/eccentric wave, 
in other words --  global disc oscillation causing a density and velocity wave
(e.g. Okazaki 1997, Lynch \& Ogilvie 2019).
The global oscillations forms a one-armed spiral density pattern 
that rotates around the star with a period of a few years (e.g. Carciofi et al. 2009). 
The variability of the emission lines of MWC~656 suggests a different mechanism 
-- pulsations induced by the orbital motion of the companion. 
Porter \& Rivinius (2003) suggested 
that in the Be/X-ray binaries, 
the neutron star has minimal (if any) impact 
on the Be star and the formation of its disc, affecting only the outer regions of the disc.
In MWC~656 the pulsations of the Be disc are visible not only 
in the outer parts of the disc (H$\alpha$ line), but also in the inner parts
(H$\beta$ and FeII lines). This is further evidence 
that the companion is more massive than the neutron stars in the Be/X-ray binaries.
The fact that the pulsations are visible in $H\beta$ and even $FeII$ lines
(see Fig.5 and Fig.6 in Zamanov et al. 2022)  
is also evidence that the orbit is eccentric.
Our results  suggest that orbital solutions in which $e=0$ is fixed and/or
$P_{orb} < 60.2$~d  can be improved.

Williams et al. (2010) pointed out that the relatively large 
Galactic latitude ($-12.3^{\circ}$) of MWC~656
suggests that the binary had been ejected 
from the Galactic plane by a supernova explosion 
that created a neutron star or black hole companion. 
Alternatively,  Dzib \& Jaron (2025)
find evidences that MWC 656 was likely formed in situ at high Galactic latitudes. 
This is one other uncertainty about MWC~656, which can be addressed 
in future studies.

Casares et al. (2014) estimate
$M_1~sin^3~i = 5.83 \pm 0.70$~M$_\odot$, $M_2~sin^3~i = 2.39 \pm 0.48 $~M$_\odot$,
$M_1 = 10 \div 12.8$~M$_\odot$ 
 and $M_2 = 3.8 \div 6.9$~M$_\odot$. 
M{\"u}ller-Horn et al.(2026) estimate  masses $M_1=7.4 \pm 2.7~M_\odot$ and 
$M_2=1.9~^{+1.7}_{-1.6}~M_\odot$. 
Our results ($M_1 = 9.0-10.6$~M$_\odot$ and $M_2 \approx 5$~M$_\odot$) agree  
with those calculated by Casares et al. (2014). 
In our Eq.~\ref{eq.3} we used the result of  M{\"u}ller-Horn et al. (2026) 
for $M_1 \sin ^3 i = 5.13 \pm 0.47$~M$_\odot$ (as given in their Table~1)  
but we obtain larger 
values for $M_1$ and  $M_2$ , 
probably due to the different approach and inclination angle used.

\section{Conclusions}

We report 53 spectra of the $H\alpha$ emission line of the Be/X-ray binary MWC~656. 
The  observations show a modulation of the EWs of the $H\alpha$ emission line 
in the range $60.29 < P_{orb} < 60.49$~d. If interpreted as an orbital period,
we find that the orbit is mildly eccentric $(e \approx 0.11)$.
We find most likely values for the components  
$M_1 = 9.0-10.6$~M$_\odot$, $M_2 \approx 5$~M$_\odot$, 
$R_1 = 9.1 \pm 0.1$~R$_\odot$,  $a \approx 160$~R$_\odot$, and $i \approx 50 - 53^\circ$. 
Our results do not prove directly that the secondary is a black hole, 
however they point that it is a massive object -- 
more massive  than the neutron stars in the Be/X-ray binaries. \\
$\; ^{ }$ \\
{\bf Data availability:} The spectra are available on ZENODO: \\
https://zenodo.org/records/22672598.

\begin{acknowledgements} 

This work is part of the project KP-06-H98/8 
"Accretion flows in binary stars" (Bulgarian National Science fund). 
The research infrastructure is supported by the National Roadmap for 
Research Infrastructure coordinated by the Ministry of Education
and Science of Bulgaria.
DM acknowledges support from Shumen University Science Fund. 
JM acknowledges support from project PID2022-136828NB-C42 funded by 
the Spanish MCIN/AEI/ 10.13039/501100011033 and "ERDF A way of making Europe".
{\ We thank an anonymous referee for making very valuable suggestions. }

\end{acknowledgements}

\clearpage 

\section*{Appendix A: Observational data}

In Table~\ref{t.1} are given the measurements of EW$_\alpha$, as described in
Sect.\ref{s.obs}. 

{\small 
\begin{table}[htb!]
\caption{The equivalent width of the $H\alpha$ emission line of MWC~656.
The first columns is date of the start of the observations, second --  the exposure time. 
HJD is the heliocentric Julian day of the mid exposure. }
\begin{center}
 \begin{tabular}{lcl ccc ccr cccll}
 \hline
 date             & exp.    & HJD (mid)          & EW$_\alpha$ & \\
                  & [min]   & 2400000+           &  [\AA]	  & \\
2021-11-11T19:20  & 	40  &      59530.32244   &	23.27  &  \\
2021-11-12T17:42  &     60  &	   59531.26140   &     23.39  &  \\
2021-11-25T20:50  &     60  &	   59544.39115   &     21.52  &  \\
2021-12-24T18:08  &     60  &	   59573.27712   &     22.68  &  \\
2022-01-19T16:52  &     45  &	   59599.21722   &     21.87  &  \\
2022-01-19T17:39  &     45  &	   59599.24949   &     21.86  &  \\
2022-01-20T17:12  &     60  &      59600.23641   &	22.33  &  \\
2022-05-13T00:11  &     60  &      59712.52601   &	23.87  &  \\
2022-05-21T22:54  &     40  &	   59721.46592   &     21.80  &  \\
2022-09-13T20:59  &     40  &	   59836.39199   &     21.20  &  \\
2023-02-10T16:58  &     60  &	   59986.22551   &     23.08  &  \\
2023-04-13T01:33  &     60  &	   60047.58200   &     22.52  &  \\
2023-07-30T00:52  &	10  &	    60155.54188   &	22.39  &  \\
2023-08-28T22:02  &	10  &	    60185.42513   &	22.19  &  \\ 
2023-08-28T22:12  &	10  &	    60185.43238   &	22.41  &  \\
2023-08-28T22:55  &	10  &	    60185.46204   &	25.36  &  \\
2023-08-28T23:06  &	10  &	    60185.46980   &	25.12  &  \\
2023-08-29T01:33  &	10  &	    60185.57186   &	22.95  &  \\
2023-08-29T01:44  &	10  &	    60185.57908   &	22.65  &  \\
2023-08-29T21:46  &	15  &	    60186.41622   &	24.79  &  \\
2023-08-29T22:02  &	15  &	    60186.42696   &	24.95  &  \\ 
2023-08-30T01:39  &	15  &	    60186.57787   &	25.14  &  \\
2023-08-30T01:55  &	15  &	    60186.58863   &	24.91  &  \\
2023-08-30T21:17  &	15  &	    60187.39611   &	25.34  &  \\
2023-08-30T21:33  &	15  &	    60187.40700   &	25.33  &  \\
2023-08-31T00:45  &	15  &	    60187.54000   &	25.55  &  \\
2023-08-31T01:00  &	15  &	    60187.55070   &	24.76  &  \\
2023-08-31T01:15  &	15  &	    60187.56141   &	25.88  &  \\ 
2023-08-31T01:31  &	15  &	    60187.57209   &	25.41  &  \\
2023-08-31T01:46  &	15  &	    60187.58280   &	25.24  &  \\
2023-08-31T02:04  &	15  &	    60187.59477   &	24.81  &  \\
2024-08-15T21:40  &	40  &	    60538.41965   &	21.19  &  \\
2024-08-21T23:43  &	10  &	    60544.49546   &	23.10  &  \\
2024-08-21T23:54  &	40  &	    60543.51349   &	21.98  &  \\
2024-08-22T22:07  &	40  &	    60545.43902   &	21.99  &  \\
2025-07-09T23:59  &	40  &	    60865.51403   &	21.92  &  \\
2025-10-29T17:49  &	60  &	    60978.26708   &	24.75  &  \\
2026-03-01T17:16  &	30  &	    61101.22679   &	25.52  &  \\
2026-03-01T17:47  &	10  &	    61101.24150   &	25.01  &  \\
2026-03-01T17:58  &	10  &	    61101.24898   &	25.32  &  \\
2026-03-01T18:09  &	10  &	    61101.25625   &	24.16  &  \\
2026-03-02T02:58  &	10  &	    61101.62369   &	24.01  &  \\
2026-03-02T03:09  &	30  &	    61101.63860   &	24.48  &  \\
2026-03-04T03:17  &	10  &	    61103.63736   &	22.11  &  \\
2026-03-04T03:29  &	30  &	    61103.65201   &	22.87  &  \\
2026-03-06T17:16  &	10  &	    61106.21958   &	23.65  &  \\
2026-04-25T01:53  &	20  &	    61155.58207   &	25.72  &  \\
2026-04-25T02:13  &	20  &	    61155.59633   &	25.73  &  \\
2026-04-26T00:56  &	20  &	    61156.54262   &	25.87  &  \\
2026-04-26T01:21  &	20  &	    61156.56000   &	26.22  &  \\
2026-04-27T01:36  &	10  &	    61157.56728   &	25.09  &  \\
2026-04-27T01:47  &	30  &	    61157.58178   &	24.48  &  \\
 \hline   								  
 \end{tabular} 							  
 \end{center}  							     
 \label{t.1}      							 			\end{table}     
}

 \begin{figure}[htb!]    
  \vspace{8.9cm}   
  \includegraphics{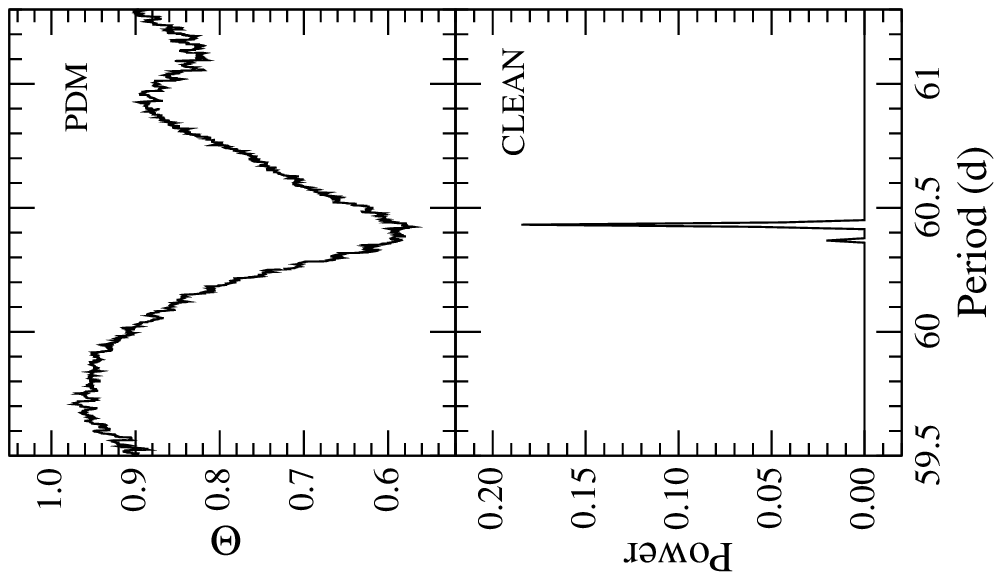}     
  \caption[]{Zoomed-in view of the results of PDM and CLEAN (see Sect~\ref{s.pa}). }
  \label{f.zoom}  
 \end{figure}

\section*{Appendix B: Uncertainties in the equations}
\label{s.a2}

About the uncertainties in the equations,
we note that: 

{\bf 1.} In Sect.~\ref{s.pa} we find orbital period, that is 1-2\% longer 
than the period found in Janssens et al. (2023)  and M{\"u}ller-Horn et al. (2026). 
This could be due to a tidal distortion   
of the Be disc (Martin et al. 2011) or Kozai-Lidov oscillations (Martin \& Franchini 2019), 
which also require significant orbital eccentricity.
We consider that the value estimated on the basis of the 
photometric and $H\alpha$ modulations
is the true orbital period, because it is 
based on a periodogram analysis and longer data sets.  
In future it will be interesting 
to see periodogram analysis of the radial velocities and/or 
to address this issue with 
different approaches 
-- X-ray, $\gamma$-ray, infrared and/or radio modulations, 
e.g. Bayesian analysis of the radio data (Gregory 1999).

{\bf 2.} Eq.~1 is based on the assumption that the FeII lines are produced in a region close 
to the surface of the Be star so the estimated value for the radius R$_1$ could be 
slightly overestimated (Hanuschik 1996, Casares et al. 2012).

{\bf 3.} Eq.~2 is used supposing that the TESS period is driven by stellar rotation. 
This period could be alternatively explained by stellar pulsations 
(Rivinius 2013, Rivinius et al. 2013).
Recent studies of the TESS light curves of Be stars show that the photometric variability is
caused by the rotation and the resulting equatorial rotational velocities are
consistent with the projected rotational velocities (Balona \& Ozuyar 2020). 
Balona \& Ozuyar (2021) suggested that 
starspots  on the surface of the star are responsible 
for the rotational modulation in B and Be stars, 
in agreement with earlier findings of Smith et al. (2006). 

{\bf 4.} We use the relation between the EW$_\alpha$ and the $R_{disc}$ as given in Eq.~4. 
If we use the relation given by Monageng et al. (2017) 
for the same values of EW$_\alpha$ the result would be
11\% larger.
As a result we would have derived larger semimajor axis and larger masses.

{\bf 5.} In the light of the results
by Okazaki \& Negueruela (2001) 
that the disc is truncated by the orbit of the companion, 
the peak of the distribution (see Fig.~3) is at $a(1-e)$. 
Following their results, it is unlikely that
the disc goes beyond $a(1+e)$, or in other words $a(1+e) \ge 176$~R$_\odot$, 
which  potentially could give a larger eccentricity and larger semi-major axis.
The cut-off of the distribution (Fig.3) is well defined, which 
suggests that $a(1+e) \approx 176$~R$_\odot$.

\end{document}